\documentclass[twocolumn,showpacs,preprintnumbers,amssymb,nofootinbib]{revtex4}

\usepackage{graphicx,epsf, epsfig, psfig, amssymb}
\usepackage{bm} 

\def\be{\begin{equation}}
\def\ee{\end{equation}}
\def\beq{\begin{eqnarray}}
\def\eeq{\end{eqnarray}}

\def\IL{\relax{\rm I\kern-.18em L}}

\begin{document}

\title{A note on the resonant frequencies of rapidly rotating black holes}

\author{Vitor Cardoso}
\email{vcardoso@teor.fis.uc.pt} \affiliation{Centro de F\'{\i}sica
Computacional, Universidade de Coimbra, P-3004-516 Coimbra,
Portugal}

\date{\today}

\begin{abstract}
I discuss the range of validity of Detweiler's formula for the
resonant frequencies of rapidly rotating Kerr black holes.  While his
formula is correct for extremal black holes, it has also been commonly
accepted that it describes very well the resonant frequencies of near
extremal black holes, and that therefore there is a large number of modes
clustering on the real axis as the black hole becomes extremal.  
I will show that this last statement is not only incorrect, but that it
also does {\it not} follow from Detweiler's formula, provided it is
handled with due care.
It turns out that only the first $n << -\log{(r_+-r_-)/r_+}$ modes are
well described by that formula, which translates, for any astrophysical
black hole, into one or two modes only.
All existing numerical data gives further support to this claim. 
I also discuss some implications of this result for recent
investigations on the late-time dynamics of rapidly rotating black
holes.

\end{abstract}


\maketitle

\section{Introduction}
Black hole spacetimes provide an interesting arena for classical wave
propagation: they absorb, scatter and sometimes even amplify incident
waves.  The detailed response of a black hole to a given incident wave
depends on the parameters characterizing the black hole, and thus
detection of a black hole (by, for example, gravitational wave
detectors) requires a deep knowledge of what precisely happens when a
wave hits a black hole.

The first numerical studies on this \cite{vish} revealed that the
response of a black hole to a given incident wavepacket is dominated
by an exponentially decaying ringing phase. What's more, the ringing
frequency and decaying timescale of this signal depends only on the
black hole parameters (mass, charge and angular momentum), and not on
the particular details of the wave packet. Thus, a black hole, like so
many other objects, has characteristic oscillation modes, called
quasinormal modes (QNMs) (the associated characteristic frequencies
are called quasinormal frequencies, QN frequencies or $\omega_{QN}$).
It turns out that these modes are excited by almost any process taking
place in the vicinities of a black hole, and thus a direct detection
of black holes may well be done by identifying this characteristic
ringing \cite{eche}.  This possibility has stimulated a great deal of
research on the computation of QNMs and QN frequencies of black holes
\cite{kokkotas} (see also \cite{cardoso1} for another recent
motivation for studying QNMs in black hole spacetimes). The situation
for Schwarzschild black holes is well understood: in Schwarzschild
coordinates, assume that the wave has the dependence 
\be
\Psi(t,r,\theta,\phi) = e^{-i(\omega t+m \phi)}Y_{lm}(\theta)R(r)\,,
\label{separation}
\ee
where $Y_{lm}(\theta)$ are just the usual spherical harmonics. I assume
$\Psi$ is a scalar wave, otherwise the separation of the angular variables
would have to be done with vectorial or tensorial spherical harmonics.
Since the Schwarzschild geometry is spherically symmetric, the azimuthal
number $m$ won't play any role.  The QN frequencies are those
frequencies $\omega$ for which the field is outgoing at infinity and
ingoing near the horizon \cite{kokkotas}. Now, it turns out that
for each angular number $l$ there is a discrete, infinite number of
frequencies satisfying these boundary conditions, and that they all
have a negative imaginary part, i.e., if we write $\omega_{QN}=\omega
_R+i\omega _I$ then $\omega _I<0$. According to the time dependence in
(\ref{separation}) the Schwarzschild geometry is then stable, since
any perturbation will decay exponentially with time.  The QN
frequencies are arranged by growing magnitude of imaginary part.  For
instance, considering gravitational waves, the fundamental mode
($n=0$) of quadrupole waves ($l=2$) is $\omega M=0.37367-0.08896$,
while the first overtone ($n=1$) is $\omega M=0.34671-0.27391$.
The fundamental mode will in general control the signal, because it is
the least damped, and it's quality factor $Q \sim \omega _R/\omega _I$ is
of order unity. Thus, we see that
Schwarzschild black holes are very poor resonators, with small $Q$'s.

The QNM spectrum for Kerr black holes is at the moment well
understood, for small rotation at least \cite{cardoso2}. 
Since this geometry is no longer
spherically symmetric, the results depend on the azimuthal number
$m$. The numerical studies indicate that for $m>0$ the quality factor
increases dramatically as one increases the rotation because the
magnitude of the imaginary part of the QN frequency decreases.
Indeed, it looks like $\omega _I \rightarrow 0$ as $a \rightarrow
M$, where $a$ is the angular momentum per unit mass and I recall that
$a$ is bounded from above by $M$; for $a=M$ we say the black hole is
extremal.  Now, what can we gain from this? The first obvious answer
is that if these modes get less damped, they should be easier to
detect.  However, this is so provided they are excited to a measurable
amplitude (see \cite{glamanderss} for a discussion on this point),
which does not happen. Still, there is an
analytical result by Detweiler \cite{detweiler} indicating that for
very rapidly rotating black holes there seems to be an infinity of QN
frequencies accumulating on the real axis. Thus, it could be that even
though a single QNM is hard to excite, all of them (which cluster on
the real axis, and are therefore nearly undamped) could conspire to give a
measurable effect.  This was investigated by Glampedakis and Andersson
\cite{glamanderss} recently, who showed that this effect is indeed
there for extremal black holes. 
They also suggest that nearly extremal
black holes could also display a similar behavior.

Here, I will discuss Detweiler's formula and its range of validity, to
see whether or not this or other phenomena involving rapidly rotating
black holes could have any importance for astrophysical black holes.
Contrary to common expectations, I find that this formula is of limited use
for any astrophysical black hole. Indeed, it describes accurately only the
first $n < -\log{(r_+-r_-)/r_+}$ overtones of rapidly rotating black holes.
For extremal black holes it should describe all the modes, and therefore
there really is an infinite number of QN frequencies clustering on the
real axis. For non-extremal black holes, 
say for example $(r_+-r_-)/r_+ \sim 10^{-10}$ (which is still
impossible to find in any astrophysical scenario \cite{thorne}) 
it describes well only a number $n << 20$ modes.

\section{The resonant frequencies of near extremal Kerr black holes}
A Kerr black hole is characterized by two parameters, the mass $M$ and the angular
momentum per unit mass $a$. It has an event horizon at $r_+=M+\sqrt{M^2-a^2}$
and a Cauchy horizon at $r_-=M-\sqrt{M^2-a^2}$.

To analyze the QNMs and QN frequencies of black holes, one studies the wave
equation that results from a linearization procedure. For Kerr black holes,
the equations describing linearized waves were derived by Teukolsky \cite{teu}.
I shall not reproduce his equations here but merely state the results, and
refer the reader to the original works \cite{teu,teupress}.
The equations describing the evolution of scalar ($s=0$), electromagnetic
($s=1$) and gravitational ($s=2$) perturbations are described by 
the following coupled equations:
\begin{eqnarray}
& & x(x+\sigma)\frac{d^2\tilde{R}_{l m}}{dx^2} \nonumber \\
& &-\{2i\hat{\omega}x^2 + 2x(2i\hat{\omega} -s-1) -(s+1)\sigma +4i\tau\}\frac{d\tilde{R}_{l m}}{dx} \nonumber \\
& &-\{2(2s+1)i\hat{\omega}(x+1) +\lambda\}\tilde{R}_{l m}=0\,, 
\label{radteuk}
\end{eqnarray}

\begin{eqnarray}
& &\frac{1}{\sin{\theta}}\frac{d}{d\theta}(\sin{\theta}\frac{dS}{d\theta})
+(a^2\omega ^2\cos{\theta}^2-\frac{m^2}{\sin{\theta}^2}-2a\omega s \cos{\theta}
\nonumber \\
& &-
\frac{2ms\cos{\theta}}{\sin{\theta}^2}-s^2\cot{\theta}^2+E-s^2)S=0\,.
\label{angteuk}
\end{eqnarray}

Here I have defined the following variables

\begin{eqnarray}
x &=& {r - r_+ \over r_+ } \ , \label{ptvar1} \\ 
\sigma &=& {r_+ - r_- \over r_+} \ , \\ 
\omega_{+} &=& \frac{a}{2Mr_+}    \, \\
\tau &=& M \left( \omega - m\omega_{+} \right) \ , \\ 
\hat{\omega} &=& \omega r_+ \, \\
\delta^2 &=& 4\hat{\omega}^2 -1/4 -\lambda \ ,
\label{PTvars}
\end{eqnarray}
where $\lambda= E + a^2 \omega^2 -2ma\omega-s(s+1)$.

These equations must be supplemented by ``physically acceptable'' 
boundary conditions. For a scattering problem, we allow in- and
out-going waves at the asymptotic region ($r \rightarrow \infty$),
but only in-going waves near the horizon, i.e., the field satisfies
\begin{equation}
\tilde{R}^{\rm in} \sim \left \{ \begin{array}{ll} 1 \quad \quad \quad 
\mbox{as} \quad r\to r_+ \ , \\   
Z^{\rm out} r^{-1} e^{2i\omega r_\ast} + Z^{\rm in} r^{-1}  
\quad \mbox{as } r\to +\infty \ .
\end{array} \right.
\label{apin}
\end{equation}
For a study of resonant frequencies, we do not want waves incident
from infinity, and thus it must be that $Z^{\rm in}=0$.
I shall also be interested in the double limit $a\to M$, 
$\omega \to m\omega_{+}$, which
corresponds to $\sigma \to 0$,
$~\tau \to 0$. 
A uniformly valid solution for (\ref{radteuk}) in this double limit
was first found by
Teukolsky and Press \cite{teupress}, following earlier work by Starobinsky
and Starobinsky and Churilov \cite{staro}.
Detweiler \cite{detweiler} then used that solution to find the QN frequencies
of near extremal black holes. I shall now re-derive the condition for the 
QN frequencies of near extremal Kerr black holes, 
making explicit the assumptions that go with it.

Let us first consider equation (\ref{radteuk}) in the limit 
when $x>>\mbox{max }(\sigma,\tau)$, i.e., for large radii. Then (\ref{radteuk}) is 
well approximated by
\begin{eqnarray}
& &x^2 \frac{d^2\tilde{R}_{l m}}{dx^2} - \{ 2i\hat{\omega}x^2 +
2x(2i\hat{\omega} -(s+1) ) \} \frac{d\tilde{R}_{l m}}{dx} \nonumber \\
& &- \{2(2s+1)i\hat{\omega}
(x+1) +\lambda \} \tilde{R}_{l m} = 0 \, .
\label{radteuk1}
\end{eqnarray}
A solution to (\ref{radteuk1}) that satisfies (\ref{apin}) can be written in 
terms of confluent hypergeometric functions \cite{stegun}
\begin{eqnarray}
\tilde{R}_{l m}&=& Ax^{-s -1/2 +2i\hat{\omega} +i\delta} M(1/2+ s +2i\hat{\omega} +i\delta, 1 +2i\delta,2i\hat{\omega} x) \nonumber \\
&+& B(\delta \rightarrow -\delta)
\label{largex}
\end{eqnarray}
where $A$,$B$ are constants and the notation $ (\delta \rightarrow -\delta)$
means ``replace $\delta$ by $-\delta$ in the preceding term''. 

We next turn to the case when $x<<1$, i.e., try to find a solution that is 
valid close to the black hole's horizon. In this regime, equation (\ref{radteuk})
can be written as
\begin{eqnarray}
\!\!\!& & x(x+\sigma)\frac{d^2\tilde{R}_{l m}}{dx^2} - \{ 2x(2i\hat{\omega} -(s+1)) 
-(s+1)\sigma + 4i\tau)\} \frac{\tilde{R}_{l m}}{dx} \nonumber \\
\!\!\!& & -\{ 2(2s+1)i\hat{\omega}+\lambda \} \tilde{R}_{l m} = 0 \ .
\end{eqnarray}
This is the hypergeometric equation, and one solution can be written as
\begin{equation}
\tilde{R}_{l m} = {}_{2}F_{1}(1/2+s -2i\hat{\omega}+i\delta, 1/2-2i\hat{\omega}+s-i\delta, 1+s- 4i\tau/\sigma,
 -x/\sigma) \ .
\label{smallx}
\end{equation} 
It is straightforward to verify that ${}_{2}F_{1}\to 1$ as $x \to 0$, 
which means that this solution has the desired ``purely ingoing wave'' 
behavior close to the event horizon. 

The key point now is that there is an overlapping region, where {\it both}
solutions are valid. Indeed, in the overlap region 
max($\sigma$,$\tau$) $\ll x \ll 1$ the solutions
(\ref{largex}) and (\ref{smallx}) both describe the solution,
and can therefore be matched.
To do the matching, take first the $x \to 0$ limit of (\ref{largex}) which yields,
\begin{equation}
\tilde{R}_{l m} \to A x^{-s-1/2 +2i\hat{\omega} +i\delta} + B(\delta \to -\delta)
\label{largex2} 
\end{equation}
Similarly, for $x \to \infty$, (\ref{smallx}) becomes,
\begin{equation} 
\tilde{R}_{l m} \to C(\delta) x^{-s-1/2 +2i\hat{\omega} +i\delta} +
C(-\delta)x^{-s-1/2 +2i\hat{\omega} -i\delta}\,,
\label{smallx2}
\end{equation} 
where
\begin{equation}
C(\delta)= \frac{ \Gamma(1 +s-4i\tau/\sigma) \Gamma(2i\delta) \sigma^{s+1/2 -2i\hat{\omega} 
-i\delta} }{ \Gamma(s+1/2 -2i\hat{\omega} +i\delta) \Gamma( 1/2 +2i\hat{\omega} +i\delta +4i\tau/\sigma) }
\label{smallxdef}
\end{equation} 
We can extract $A$ and $B$ by matching the solutions (\ref{largex2}) and
(\ref{smallx2}), 
\begin{eqnarray}
A &=& C(\delta)\frac{\Gamma(1/2 +2i\hat{\omega} + i\delta +4i\tau/\sigma)}
{\Gamma(1/2 +2i\hat{\omega} + i\delta -4i\tau/\sigma) }
\label{A}
\\
B &=& C(-\delta)\frac{\Gamma(1/2 +2i\hat{\omega} + i\delta +4i\tau/\sigma)}
{\Gamma(1/2 +2i\hat{\omega} + i\delta -4i\tau/\sigma) }
\label{B}
\end{eqnarray}
On the other hand, approximating (\ref{largex}) for $x \to \infty$ we get for the amplitudes $ Z^{\rm in}$,$Z^{\rm out}$,
\begin{eqnarray}
Z^{\rm in}&=& A \frac{\Gamma(1+2i\delta)}{\Gamma(1/2-s -2i\hat{\omega} +i\delta)}
(-2i\hat{\omega})^{-1/2-s -2i\hat{\omega} -i\delta} \nonumber \\
&+& B(\delta \to -\delta) 
\end{eqnarray}

\begin{eqnarray}
Z^{\rm out} &=& A \frac{\Gamma(1+2i\delta)}{\Gamma(1/2+s +2i\hat{\omega} + i\delta)}
(2i\hat{\omega})^{-1/2+s +2i\hat{\omega} -i\delta} \nonumber \\
&+& B(\delta \to -\delta)
\end{eqnarray}

To find the resonant frequencies we now impose $Z^{\rm in}=0$.
Using (\ref{A})-(\ref{B}) we find that for these frequencies the following
condition must be satisfied
\begin{eqnarray}
& &- { \Gamma(2i\delta) \Gamma(1+2i\delta)  \over \Gamma(-2i\delta)
\Gamma(1-2i\delta)}\left[{\Gamma(s+1/2-2i\hat{\omega}-i\delta) \over
\Gamma(s+1/2-2i\hat{\omega}+i\delta) } \right]^2 \nonumber \\
& &=(-2i\hat{\omega}\sigma)^{2i\delta} 
{\Gamma(1/2+2i\hat{\omega}+i\delta-4i\tau/\sigma) \over
\Gamma(1/2+2i\hat{\omega}-i\delta-4i\tau/\sigma) } \ .
\label{modecond}
\end{eqnarray}

\subsection{Detweiler's formula}
Up to this point, the only assumptions that entered the derivation
of (\ref{modecond}) were that the black hole is very rapidly spinning
and that the $\omega \sim m\omega _+$. 
Let us now proceed to derive Detweiler's formula for the QN frequencies
of rapidly rotating black holes.

The quantity $\delta$ is in general a complex quantity. However,
the numerical studies show that for $\omega \sim m \omega _+$, and for 
{\it some} $l,m$ values it is almost purely real. I would like to
emphasize that this fact is seen numerically, it hasn't been analytically
proved. Moreover, it seems to be true for most $l=m$ modes but not only:
$\delta$ is real also for some other values of $l,m$, and so there is 
nothing special about $l=m$ modes, contrary to what is stated in \cite{glamanderss}.
To be more precise, I show in Table \ref{tab:delta} the values of $\delta ^2$ for
some $l,m$ pairs, assuming one is exactly in the extreme case, i.e., $a\omega=m/2$.
The Table refers to the scalar case $s=0$, although the trend is similar for other
$s$ values.
Notice that for $l=m=1$ the quantity $\delta ^2$ is negative, and notice also
that for $l=6\,,m=5$ it is positive. So, there really is nothing special
about $l=m$ modes.
\begin{table}
\centering
\caption{\label{tab:delta} 
Values of the constant $\delta ^2$ for some ($l, m$) pairs, for
$a\omega=m/2$, i.e., the extreme limit, and for scalar
perturbations. Notice two important things:
first, for $l=m=1$ the quantity $\delta ^2$ is negative, and thus
$\delta$ is purely imaginary. Second, for $l=6\,,m=5$, $\delta$
is purely real. Thus, although the rule is that for $l=m$ the quantity
$\delta$ is real, while for other $l\,,m$ values it is imaginary,
there are exceptions. 
}
\vskip 12pt
\begin{tabular}{@{}ccc|ccc@{}}  
\hline
\hline
$l$ &$m$  &$\delta ^2$  & $l$ &$m$  &$\delta ^2$ \\
\hline
\hline            
1 & 1 &-0.4497   &    3 & 2 & -4.9144   \\
2 & 2 & 0.8948   &    4 & 4 &  8.1233   \\
2 & 1 &-4.3926   &    4 & 3 & -3.8792   \\
3 & 3 & 3.7552   &    6 & 5 &  2.7752  \\  
\hline 
\hline
\end{tabular}
\end{table}

With this cleared, I shall go on assuming that $\delta$ is a real quantity.

The left-hand side of equation (\ref{modecond}) has a well-defined limit as
$a\to M$ and $\omega \to m\omega_{+}$. We represent that limit by
\begin{equation}
\mbox{LHS } = qe^{i\chi} \ .
\end{equation}

Now, we cannot have a consistent solution unless
$\tau/\sigma\to \infty$ as $a\to M$; this is because if $\tau/\sigma$
had some finite limit, then so would the Gamma functions in the RHS of
equation (\ref{modecond}). And then we would have 
$\hat{\omega}\sim \frac{X}{\sigma}$, with $X$ a well defined quantity.
But this would violate our assumptions that $\omega \sim m\omega _+$.
Therefore, let us proceed assuming that $\tau/\sigma\to \infty$.
Using Stirling's formula \cite{stegun},
\be
\Gamma[b+az]\sim \sqrt{2\pi}e^{-az}(az)^{b+az-1/2}\,,
\label{Stir}
\ee
the right-hand side of the (\ref{modecond}) can be written
\begin{equation}
\mbox{ RHS } = (-8\hat{\omega}\tau)^{2i\delta} \ .
\end{equation}
In other words, a QNM must be a solution to
\begin{equation}
f(\omega) = (-8\hat{\omega}\tau)^{2i\delta} - qe^{i\chi} = 0 \ .
\end{equation}
Using $-8\hat{\omega}\tau = \rho e^{i\zeta}$ we see that
solutions follow from (remembering that $\delta$ is real and $\omega \to \omega_+$)
\begin{equation}
\rho = \exp\left[ { \chi - 2k\pi \over 2\delta} \right]\,,
\end{equation}
\begin{equation}
\zeta = - {1\over 2\delta} \log{q} \,,
\end{equation}
where $k$ is any integer number.
This is Detweiler's formula \cite{detweiler}. Since we are always assuming
$\omega \approx m\omega _+$, this can also be written as \cite{sasaki}
(just substitute $-8\hat{\omega} \tau \sim -4m(\omega-m/2)$)
\begin{equation}
\omega M \approx {m\over 2} - {1\over 4m}e^{(\chi - 2k\pi)/
2\delta} \cos \zeta - {i\over 4m}e^{(\chi - 2k\pi)/ 2\delta}
\sin \zeta \ .
\label{llmodes}
\end{equation}
It is possible to prove \cite{sasaki} that $\sin\zeta >0$, i.e. these QNMs 
are all damped. 

\subsection{Discussion of Detweiler's formula}
I now turn to discuss what is the range of validity of formula (\ref{llmodes}).
We have been assuming that $\tau/\sigma\to \infty$, but there is a number $k$
in (\ref{llmodes}) at which this condition breaks down.
In fact, $\tau >> \sigma$ implies, according to (\ref{llmodes}) that
in order of magnitude one should have
\be
e^{-k}>>\frac{r_+-r_-}{r_+}\,.
\label{cond11}
\ee
Thus, there is an upper bound on $k$ given by
\be
k<<-\log{\frac{r_+-r_-}{r_+}}\,.
\label{cond12}
\ee
Notice that formula (\ref{cond12}) is basically a lower bound for the fundamental
mode. In fact, with (\ref{cond12}) we get, from equation (\ref{llmodes}) that
$\omega _I M>>(r_+-r_-)/r_+$.

This is a very restrictive bound. For instance, if we have
$(r_+-r_-)/r_+ \sim 10^{-10}$, formula (\ref{llmodes}) can only describe
well the first $n << 20$ modes, where $n$ now labels the overtone number.  
We know that accretion cannot spin a black hole
beyond $a=0.998M$ \cite{thorne}, and so, for any astrophysical
black hole, Detweiler's formula should only describe accurately one or two modes.

Extremal black holes are accurately described by (\ref{llmodes}) for all $n$.
In fact, all one has to do to prove this is to put $\sigma=0$ in 
Teukolsky's equations. In this case equation (\ref{smallx}) would be replaced by
a confluent hypergeometric, and the final outcome for the QN frequencies 
would still be equation (\ref{llmodes}). Thus, for extremal black holes, there is 
indeed an infinity of modes clustering on the real axis.
\section{Discussion and implications}
I have re-derived Detweiler's \cite{detweiler} result for the quasinormal 
frequencies of rapidly rotating black holes, and I have shown to what extent 
it can be used.
I have proved that for any astrophysical black hole, Detweiler's
approximation is likely to be accurate for just the fundamental
mode. 
In \cite{glamanderss} it was shown that the signal from extremal black
holes goes like $t^{-1}$ at late times instead of the usual power-law
$t^{-2l-3}$ (for $l=m$, but I refer the reader to \cite{hod}) .
Although they prove this for $l=m$ modes there should be other modes
decaying like this, according to the discussion in this paper.  Now,
the main ingredient in their result was the fact that for extremal
black holes there is an infinity of long-lived modes near the real
axis.  What hope is there to have this same phenomenon for
non-extremal, but rapidly rotating black holes?  As I have shown, this
effect will not be present for any astrophysical realistic black
hole, not according to the analytical approximation, and not according
to the numerical results in \cite{cardoso2}.
To conclude, I would like to draw attention to the fact that Detweiler's
approximation holds good for any ($l\,,m$) pair for which the quantity
$\delta $ is real. This includes most of the $l=m$ modes, and some other
modes, but not all. However, from the numerical work in \cite{cardoso2}
it seems that {\it all} modes having positive $m$ tend to cluster
on the real axis, as the black hole approaches extremality. One is thus
led to suspect that there should be some general argument, independent 
of $\delta$, showing that all modes cluster on the real axis. 
\section*{Acknowledgements}
I am grateful to Emanuele Berti for a critical 
reading of the manuscript, and for many useful suggestions.
I would also like to thank Kostas Glampedakis for useful correspondence
on this problem, and acknowledge financial support from FCT
through grant SFRH/BPD/2003.


\begin{thebibliography}{99}

\bibitem{vish} C. V. Vishveshwara, 
Nature {\bf 227}, 936 (1969).

\bibitem{eche} F. Echeverria,
Phys. Rev. D {\bf 40}, 3194 (1989);
L. S. Finn, 
Phys. Rev. D {\bf 46}, 5236 (1992);
H. Nakano, H. Takahashi, H. Tagoshi, M. Sasaki,
gr-qc/0306082;
O. Dreyer, B. Kelly, B. Krishnan, L. S. Finn, D. Garrison, R. Lopez-Aleman,
Class. Quant. Grav. {\bf 21}, 787 (2004).


\bibitem{kokkotas} K.D. Kokkotas and B.G. Schmidt,
Living Rev. Relativ. {\bf 2}, 2 (1999);
H.-P. Nollert, 
Class. Quant. Grav. {\bf 16}, R159 (1999).

\bibitem{cardoso1}  V. Cardoso, J. P. S. Lemos and S. Yoshida,
Phys. Rev. D {\bf 69}, 044004 (2004). 

\bibitem{cardoso2} E. W. Leaver, 
Proc. R. Soc. London {\bf A402}, 285 (1985).
H. Onozawa,
Phys. Rev. D {\bf 55}, 3593 (2003);
E. Berti and K. D. Kokkotas, 
Phys. Rev. D {\bf 68}, 044027 (2003); 
E. Berti, V. Cardoso, K. D. Kokkotas and H. Onozawa,
Phys. Rev. D {\bf 68}, 124018 (2003);
E. Berti, V. Cardoso and S. Yoshida,
Phys. Rev. D {\bf 69}, 124018 (2004);
S. Musiri and G. Siopsis, 
Phys. Lett. B {\bf 579}, 25 (2004);
For a recent overview on this subject see
E. Berti, gr-qc/0411025.

\bibitem{glamanderss} K. Glampedakis and N. Andersson,
Phys. Rev. D {\bf 64}, 104021 (2001).

\bibitem{detweiler} S. Detweiler,
Astrophys. J. {\bf 239}, 292 (1980).

\bibitem{thorne} K. S. Thorne,
Astrophys. J. {\bf 191}, 507 (1974).

\bibitem{teu} S. A. Teukolsky,
Astrophys. J. {\bf 185}, 635 (1973).

\bibitem{teupress} S. A. Teukolsky and W. H. Press,
Astrophys. J. {\bf 193}, 443 (1974).

\bibitem{staro} A. A. Starobinsky,
Zh. Eksp. Teor. Fiz {\bf 64}, 48 (1973)
[Sov. Phys. JETP {\bf 37}, 28 (1973)];
A. A. Starobinsky and S. M. Churilov,
Zh. Eksp. Teor. Fiz {\bf 65}, 3 (1973)
[Sov. Phys. JETP {\bf 38}, 1 (1973)].

\bibitem{stegun} M. Abramowitz, I. A. Stegun, 
{\it Handbook of Mathematical Functions},
(Dover, New York, 1970). 

\bibitem{sasaki} M. Sasaki and T. Nakamura, 
Gen. Rel. Grav. {\bf 22}, 1351 (1990).

\bibitem{hod} S. Hod,
Phys. Rev. D {\bf 58}, 104022 (1998);
Phys. Rev. D {\bf 60}, 104053 (1999);
Phys. Rev. D {\bf 61}, 024033 (2000);
Phys. Rev. D {\bf 61}, 064018 (2000).

\end{thebibliography}
\end{document}